   \documentclass[twocolumn,showpacs,preprintnumbers,amsmath,amssymb]{revtex4}

   \usepackage{psfig} 
   \usepackage{graphicx} 
\usepackage{amsfonts}
\usepackage{amsmath}
\usepackage{amssymb}
\usepackage{tabularx}
\usepackage{subfigure}
\usepackage{psfrag}
\usepackage{url}

\begin{document}

 \title{Metrics for more than two points at once}

   \author{David H. Wolpert$^1$}
   \affiliation{$^1$NASA Ames Research Center, Moffett Field, CA 94035,
    \tt \{dhw@email.arc.nasa.gov}

   \begin{abstract} 
The conventional definition of a topological metric over a
space specifies properties that must be obeyed by any measure of ``how
separated" two points in that space are. Here it is shown how to
extend that definition, and in particular the triangle inequality, to
concern arbitrary numbers of points. Such a measure of how
separated the points within a collection are can be bootstrapped, to
measure ``how separated" from each other are two (or more)
collections. The measure presented here also allows fractional
membership of an element in a collection. This means it directly
concerns measures of ``how spread out" a probability distribution over
a space is. When such a measure is bootstrapped to compare two
collections, it allows us to measure how separated two probability
distributions are, or more generally, how separated a distribution of
distributions is.
   \end{abstract}

\maketitle

 \section{Introduction}

The conventional definition of a topological metric formalizes the
concept of distance. It specifies properties required of any function
that purports to measure ``how separated" two elements of a space
are. However often one wants to measure ``how separated" the members
of a collection of more than two elements is. The conventional way to
do this is to combine the pair-wise metric values for all pairs of
elements in the collection, into an aggregate measure. This is ad hoc
however.

As an alternative, here the formal definition of a topological metric
is extended to apply to collections of more than two elements.  In
particular, the triangle inequality is extended to concern such
collections.  The measure presented here applies even to collections
with duplicate elements. It also applies to collections with
``fractional" numbers of elements, i.e., to probability distributions.

This measure can be directly incorporated into many domains where ad
hoc combinations of pair-wise metrics are currently used. In addition,
when applied to different projections of a high-dimensional data set,
it provides a novel type of vector-valued characterization of that
data set.

This new measure can be bootstrapped in a natural way, to measure
``how separated" from each other two collections are. In other words,
given a measure $\rho$ of how separated from each other the elements
in an arbitrary collection $\xi$ are, one can define a measure of how
separated from each other two collections $\xi_1$ and $\xi_2$ are.
(Intuitively, the idea is to subtract the sum of the measure's values
for each of the two separate collections $\xi_1$ and $\xi_2$ from the
value of the measure for the union of the collections.) More
generally, one can measure how separated a collection of such
collections is. Indeed, with fractional memberships, such
bootstrapping allows us to measure how separated a distribution of
distributions is.

In the next section the definition of a multi-argument metric
({\bf{multimetric}}, for short) is presented. Also in that section is
an extensive set of examples and a list of some elementary
properties. For instance, it is shown that the standard deviation of a
probability distribution across $\mathbb{R}^N$ is a multimetric,
whereas the variance of that distribution is not.

The following section presents a way to bootstrap from a multimetric
for elements within a collection to a multimetric over collections.
Some examples and elementary properties of this bootstrapped measure
are also in that section.

A short concluding section considers some of the possible uses of
multimetrics.

 \section{Multimetrics}

Collections of elements from a space $X$ are represented as vectors of
counts, i.e., functions from $x \in X \rightarrow \{0, 1, 2,
\ldots\}$.  So for example, if $X = \{A, B, C\}$, and we have the
collection of three $A$'s, no $B$'s, and one $C$, we represent that as
the vector $(3, 0, 1)$. It is natural to extend this to functions from
$x \in X$ to $\mathbb{R}$.  In particular, doing this will allow us to
represent probability distributions (or density functions, depending
on the cardinality of $X$) over $X$. Accordingly, our formalization of
multimetrics will provide a measure for how how spread out a
distribution over distributions is.\footnote{See
\cite{lin91,hall99,judg04,osva03,futo04,woma04a} and references
therein for work on how spread out a pair of distributions is.}  Given
$X$, the associated space of all functions from $X$ to $\mathbb{R}$ is
written as $\mathbb{R}^X$. The subspace of functions that are
nowhere-negative is written as $(\mathbb{R}^+)^X$.

As a notational comment, integrals are written with the measure
implicitly set by the associated space. In particular, for a finite
space, the point-mass measure is implied, and the integral symbol
indicates a sum. In addition $\delta_{x}$ is used to indicate the
appropriate type of delta function (Dirac, Kronecker, etc.) about $x$.
Other shorthand is ${\cal{R}}^X \equiv ({\mathbb{R}}^+)^X -
\{0\}$ and $||v||$ to mean $\int dx \;v(x)$.

In this representation of collections of elements from $X$, any
conventional metric taking two arguments in $X$ can be written as a
function $\rho$ over a subset of the vectors in ${\cal{R}}^X$. That
subset consists of all vectors that either have two of their
components equal to 1 and all others 0, or one component equal to 2
and all others 0. For example, for $X = \{A, B, C\}$, the metric
distance between $A$ and $B$ is $\rho(1,0,1)$, and from $A$ to itself
is $\rho(2,0,0)$.

Generalizing this, a multimetric for $T(X) \subseteq {\cal{R}}^X$ is
defined as a real-valued function $\rho$ over ${\cal{R}}^X$ such that
$\forall u, v, w \in {\cal{R}}^X$,

$ $

1) $u,v,w \in T(X) \;\;\Rightarrow\;\; \rho(u + v) \; \le \; \rho(u +
w) + \rho(v + w).$

$ $

2) $\rho(u) \ge 0, \; \rho(k \delta_{x}) = 0 \; \forall x, k > 0.$

$ $

3) $\rho(u) = 0  \;\;\Rightarrow \;\; u = k \delta_{x} \; {\mbox{for
    some}} \; k, x.$

$ $

In this representation of collections, if only one $x \in X$ is in a
collection (perhaps occurring more than once), then only one component
of $u$ is non-zero. Accordingly, conditions (2) and (3) are extensions
of the usual condition defining a metric that it be non-negative and
equal 0 iff its arguments are the same. Condition (1) is an extension
of the triangle inequality, to both allow repeats of elements from $X$
and/or more than two elements from $X$ to be in the collection. Note
though that condition (1) involves sums in its argument rather than
(as in a conventional norm-based metric for a Euclidean space)
differences. Intuitively, $T(X)$ is that subset of ${\cal{R}}^X$ over
which the generalized version of the triangle inequality
holds.\footnote{A natural extension of this analysis allows
${\cal{R}}^X$ to be any subset of ${\mathbb{R}}^X$ --- potentially
including vectors with negative components --- so long as $k \delta_x
\in {\cal{R}}^X \; \forall x \in X, k > 0$. Indeed, with care, one can
even take the range of $\rho$ to include complex numbers, so that
$\rho$ includes Hilbert vectors like those that arise in quantum
mechanics. In the interests of clarity, we do not pursue such
extensions here.}

Condition (1) implies that multimetrics obey a second triangle
inequality, just as conventional metrics do:

$ $

$\rho(u + v) \ge |\rho(u + w) - \rho(v + w)|$.

$ $

 \noindent (This follows by rewriting condition (1) as $\rho(u + w) \ge
\rho(u + v) - \rho(v + w)$, and then relabeling twice.) 

$ $

 \noindent {\bf{Example 1}}: Set $X = \mathbb{R}^N$. Take $T(X)$ to be those
elements of ${\cal{R}}^X$ whose norm equals 1, i.e., the probability
density functions over $\mathbb{R}^N$.  Then have $\rho(s)$ for any $s
\in {\cal{R}}^X$ (whether in $T(X)$ or not) be the standard
deviation of the distribution $\frac{s}{||s||}$, i.e., $\rho(s) =
\sqrt{\frac{1}{2}\int dx dx' \; \frac{s(x) s(x')}{||s||^2} (x -
x')^2}$.

Conditions (2) and (3) are immediate. To understand condition (1),
first, as an example, say that all three of $u, v$ and $w$ are
separate single delta functions over $X$. Then condition (1) reduces
to the conventional triangle inequality over $\mathbb{R}^N$, here
relating the points (in the supports of) $u, v$ and $w$. This example
also demonstrates that the variance (i.e., the square of our $\rho$)
is not a multimetric.

For a vector $s$ that involves multiple delta functions, $\rho(s)$
measures the square root of the sum of the squares of the Euclidean
distances between the points (in the support of) $s$. In this sense it
tells us how ``spread out'' those points are. Condition (1) even holds
for vectors that are not sums of delta functions however (see appendix). 

$ $

 \noindent {\bf{Example 2}}: As a variant of Ex. 1, have $X$ be the
unit simplex in $\mathbb{R}^N$, and use the same $\rho$ as in
Ex. 1. In this case any element of $X$ is a probability distribution
over a variable with $N$ possible values. So any element of $T(X)$ is
a probability density function over such probability distributions. In
particular, say $s$ is a sum of some delta functions for such an
$X$. Then $\rho(s)$ measures how spread out the probability
distributions in (the support of) $s$ are. If those probability
distributions are themselves sums of delta functions, they just
constitute subsets of our $N$ values, and $\rho(s)$ measures how
spread out from one another those subsets are.

$ $

 \noindent {\bf{Example 3}}: As another variant of Ex. 1, for any
$X$, take $T(X) = {\cal{R}}^X$. Define the tensor contraction $\langle
s \mid t \rangle \equiv \int dx dx' s(x) t(x) F(x, x')$ where $F$ is
symmetric and nowhere-negative, and where $F(x, x') = 0
\Leftrightarrow x = x'$. Then $\rho(s) \equiv \sqrt{{\langle}s \mid
s{\rangle}}$ obeys conditions (2) and (3) by inspection.  It also
obeys condition (1) (see appendix).

Note that the ${\langle} ., . \rangle$ operator is not an inner
product over ${\mathbb{R}}^X$, the extension of $T(X)$ to a full vector
space. When components of $s$ can be negative, $\langle s, s \rangle$
may be as well.  Note also that there is a natural differential
geometric interpretation of this $\rho$ when $X$ consists of $N$
values. Say we have a curve on an $N$-dimensional manifold with metric
tensor $F$ at a particular point on the curve, and that at that point
the tangent vector to the curve is $s$. Then $\rho(s)$ is the
derivative of arc length along that curve, evaluated at that point.

This suggest an extension of this multimetric, in which rather than a
tensor contraction between two vectors, we form the tensor contraction
of $n$ vectors: ${\langle}s^1, \ldots , s^n{\rangle} \equiv \int dx^1
\ldots dx^n s^1(x^1) \ldots s^n(x^n) F(x^1, \ldots, x^n)$, where $F$
is invariant under permutation of its arguments, nowhere-negative, and
equals 0 if and only if all its arguments have the same value. Any
$\rho(s)$ that is a monotonically increasing function of ${\langle}s,
s, \ldots, s{\rangle}^{1/n}$ automatically obeys conditions (2) and
(3).

$ $

It is worth collecting a few elementary results concerning
multimetrics:

$ $

 \noindent {\bf{Lemma 1}}:

 \begin{enumerate}

 \item Let $\{\rho_i\}$ be a set of functions that obey conditions (2)
and (3), and $\{a_i\}$ a set of non-negative real numbers at least one
of which is non-zero. Then $\sum_i a_i \rho_i$ also obeys conditions
(2) and (3).

 \item Let $\{\rho_i\}$ be a set of functions that obey condition (1),
and $\{a_i\}$ a set of non-negative real numbers at least one of which
is non-zero. Then $\sum_i a_i \rho_i$ also obeys conditions (1).

 \item Let $f : {\mathbb{R}} \rightarrow {\mathbb{R}}^+$ be a
monotonically increasing concave function that equals 0 when its
argument does.  Then if $\rho$ is a multimetric for some $T(X)$,
$f(\rho)$ is also a multimetric for $T(X)$ (see appendix).

 \item Let $f : X \rightarrow Y$ be invertible, and let $\rho_Y$ be a
multimetric over $Y$. Define the operator $B_f : {\cal{R}}^X
\rightarrow {\cal{R}}^Y$ by $[B_f(s)](y) \equiv s(f^{-1}(y))$ if
$f^{-1}(y)$ exists, 0 otherwise. $B_f$ is a linear operator. This
means $\rho_X(s) \equiv \rho_Y(A_f(s))$ is a
multimetric.\footnote{Such mappings are analogous to the mapping from
a ``data space" to a ``feature space" underlying kernel machines
~\cite{chsh00}.}

 \end{enumerate}

$ $

 \noindent {\bf{Example 4}}: Take $X = \mathbb{R}^N$ again, and let
$T(X)$ be all of ${\cal{R}}^X$ with bounded support.  Then by Lemma 1,
the width along $x_1$ of (the support of) $s \in T(X)$ is a
multimetric function of $s$ (see appendix).

This means that the average of the width in $x_1$ over all possible
rotations of $X$ is also a multimetric. Similarly, consider the
smallest axis-parallel box enclosing the (support of the) Euclidean
points in $s$.  Then the sum of the lengths of the edges of that box
is a multimetric function of $s$.  

On the other hand, while the volume of that box obeys conditions (2)
and (3), in general it can violate condition (1). Similarly, the
volume of the convex hull of the (support of) the points in $s$ obeys
conditions (2) and (3) but can violate (1). (In general, multimetrics
have the dimension of a length, so volumes have to be raised to the
appropriate power to make them be multimetrics.)

It is worth comparing the sum-of-edge-lengths multimetric to the
standard deviation multimetric of Ex. 1 for the case where all
arguments $s$ are finite sums of delta functions (i.e., ``consist of a
finite number of points''). For such an $s$ we can write the
sum-of-edge-lengths multimetric as a sum over all $N$ dimensions $i$
of ${\mbox{max}}_j \; s^j_i - {\mbox{min}}_j \; s^j_i$, where $s^j$ is
the $j$'th point in $s$. In contrast, the (square of the) standard
deviation multimetric is also a sum over all $i$, but of the (square
of the) standard deviation of the $i$'th components of the points in
$s$. Another difference is that the standard deviation multimetric is
a continuous function of its argument, unlike the sum-of-edge-lengths
multimetric.

$ $

 \noindent {\bf{Example 5}}: Let $X$ be countable and have $T(X) =
{\cal{R}}^X$. Then $\rho(s) = \int dx \Theta(s(x)) - 1$ where $\Theta$
is the Heaviside function is a multimetric (see appendix). This is the
volume of the support of $s$, minus 1.

$ $

 \noindent {\bf{Example 6}}: Let $X$ be countable and have $T(X) =
{\cal{R}}^X$. Then $\rho(s) = ||s|| - {\mbox{max}}_x s(x)$ obeys
conditions (2) and (3), by inspection. Canceling terms, for this
$\rho$ condition (1) holds iff ${\mbox{max}}_x (u(x) + v(x)) \ge
{\mbox{max}}_x (u(x) + w(x)) + {\mbox{max}}_x (v(x) + w(x)) -
2||w||$. This is not true in general, for example when $||w|| = 0$ and
the supports of $u$ and $v$ are disjoint. However if we take $T(X)$ to
be the unit simplex in ${\cal{R}}^X$, then condition (1) is obeyed,
and $\rho$ is a multimetric (see appendix).

$ $

 \noindent {\bf{Example 7}}: Let $X$ have a finite number of elements and
set $T(X) = {\cal{R}}^X$. Say that $\rho(s) = 0$ along all of the
axes, and that everywhere else, $k \le \rho(s) \le 2k$ for some fixed
$k > 0$. Then $\rho$ is a multimetric.

 \subsection{Vector-valued multimetrics}

It is straightforward to extend the definition of a multimetric to
have range ${\mathbb{R}}^M$ rather than $\mathbb{R}$, so long as one
has a linear ordering over ${\mathbb{R}}^M$ to specify the appropriate
extension of condition (1). For example, consider the component-wise
ordering: $\forall a, b, \vec{a} \le \vec{b} \Leftrightarrow a_i \le
b_i \; \forall i \in \{1, 2, \ldots M\}$. Say we have a set of $M$
scalar multimetrics. Then the $M$-fold Cartesian product of those
multimetrics is an $M$-dimensional multimetric, when component-wise
ordering defines the inequality in condition (1).

More generally, say we have chosen such a linear ordering over
${\mathbb{R}}^M$, and have an $M$-dimensional function with domain
${\cal{R}}^X$. Say this function obeys conditions (1) and (2) of an
$M$-dimensional multimetric for our linear ordering.  Then this
function can be used as a low-dimensional characterization of an
element of ${\cal{R}}^X$. In general, such characterizations may have
$M$ is less than $|{\mathbb{R}}^X|$, the dimension of the space in
which ${\cal{R}}^X$ is embedded, and may violate condition (1).  The
following examples illustrates this:

$ $

{ \noindent {\bf{Example 8}}: Consider again Ex. 1. To define our
vector-valued multimetric for the  $X$ of Ex. 1, say we have a scalar
multimetric $\rho_{\mathbb{R}}$ for the subspace, $X' = T(X') =
{\mathbb{R}}$. Let $\{v^1, v^2, \ldots v^M\}$ be a set of $M$
unit norm vectors living in $X$. Then we can define our
$M$-dimensional multimetric by
 \begin{eqnarray*} 
\rho_i(u) &\equiv& \rho_{\mathbb{R}}[f_{u, v^i}(t)] ; \\
f_{u, v^i}(t) &\equiv& \int dx \; u(x) \delta(v^i \cdot x - t).
 \end{eqnarray*}

To illustrate this, take $M= N$ and have the $\{v^i\}$ be the unit
normals along the $N$ axes of $X$. Let $u$ be a sum of delta
functions; $u = \delta_{x^1} + \delta_{x^2}$. Let $\rho_{\mathbb{R}}$
be the standard deviation multimetric of Ex. 1 for one-dimensional
probability density functions. So each component $\rho_i(u)$ is just
the $i$'th component of the difference $x^1 - x^2$.  Accordingly, $u$
can be reconstructed from the vector $\rho_i(u)$.

In this illustration conditions (1) and (2) are immediate.  If $M =
N$, then condition (3) also holds for $u$'s like the one considered
here that are sums of two delta functions, but not more generally.
Now modify this illustration by having $M < N$ and the $\{v^i\}$ not
all point along the axes of $X$. Then for general $u$, the components
$\rho_i(u)$ are the projections of $u$ along the different vectors
$\{v^i\}$. As in techniques like Principal Components
Analysis~\cite{duha00}, those projections provide a low-dimensional
characterization of $u$.

 \section{Concavity gaps and dispersions}

In Ex. 1, $\rho$ can be used to tell us how spread out a distribution
over $\mathbb{R}^N$ is. One would like to be able to use that $\rho$
to construct a measure of how spread out a collection of multiple
distributions over $\mathbb{R}^N$ is. Intuitively, we want a way to
construct a metric for a space of sets (generalized to be able to work
sets with duplicates, fractional memberships, etc.) from a metric for
the members within a single set. This would allow us to directly
incorporate the distance relation governing $X$ into the distance
relation for ${\cal{R}}^X$.

To do this, first let $\{Y, S(Y)\}$ be any pair of a subset of a
vector space together with a subset of $\mathbb{R}^Y$ such that
$\forall g \in S(Y), \frac{\int dy g(y) y}{||g||} \in Y$. (As an
example, we could take $Y$ to be any convex subspace of a vector
space, with $S(Y)$ any subset of ${\cal{R}}^Y$.) Then the associated
{\bf{concavity gap}} operator ${\cal{C}} : S(Y) \rightarrow
{\mathbb{R}}^{S(Y)}$ is defined by
 \begin{eqnarray*}
({\cal{C}}\sigma)(g) = \sigma(\frac{\int dy \; g(y) y}{||g||}) -
\frac{\int dy \; g(y) \sigma(y)}{||g||}
 \end{eqnarray*}
 \noindent where $y \in Y$, and both $\sigma$ and $g$ are arbitrary
elements in $S(Y)$.  So the concavity gap operator takes any single
member of the space $S(Y)$ (namely $\sigma$) and uses it to generate a
function (namely, ${\cal{C}}\sigma$) over all of
$S(Y)$.\footnote{Equivalently, it can be viewed as a non-symmetric
function from $S(Y) \times S(Y) \rightarrow
\mathbb{R}$, although we will not exploit that perspective here.}

In particular, say $Y = T(X)$ for some space $X$. Say we are given a
multimetric $\sigma$ measuring the ($X$-space) spread specified by any
element of $Y$. Say we are also given a $g$ which is a normalized
distribution over $Y$. Then ${\cal{C}}\sigma(g)$ is a measure of how
spread out the distribution $g$ is. Note that in this example space
$S(Y)$ is both the space of multimetrics over $Y$ and the space of
distributions for $Y$, exemplified by $\sigma$ and $g$, respectively.

We can rewrite the definition of the concavity gap in several ways:
 \begin{eqnarray*}
{\cal{C}}\sigma(g) &=& \sigma(E_g(y)) - E_g(\sigma) \\
&=& \sigma(\frac{{\bf{y}} \cdot g}{||g||}) - \frac{\sigma \cdot
g}{||g||}
 \end{eqnarray*}
 \noindent where $E_g$ means expected value evaluated according to the
probability distribution $\frac{g}{||g||}$, and in the last expression
{\bf{y}} is the (infinite-dimensional) matrix whose $y$'th column is
just the vector $y$, and the inner products are over the vector space
$S(Y)$. Taken together, these equations say that the concavity gap of
$\sigma$, applied to the distribution $g$, is given by evaluating the
function $\sigma$ at the center of mass of the distribution $g$,
and then subtracting the inner product between $\sigma$ and $g$.

$ $

 \noindent {\bf{Example 9}}: Let $Y = {\mathbb{R}}^N$, and choose
$S(Y)$ to be the set of nowhere-negative functions of $Y$ with
non-zero magnitude. Choose $\sigma(y) = 1 - \sum_{i=1}^N y_i^2$. Then
${\cal{C}}\sigma(g) = Var(\frac{g}{||g||})$.

$ $

 \noindent {\bf{Example 10}}: Say $X$ has $N$ values, with $T(X) =
{\cal{R}}^X$. Consider a $u \in T(X)$ whose components are all either
0 or some particular constant $a$ such that $\int dx \; u(x) = 1$. So
$u$ is a point on the unit hypercube in $T(X)$, projected down to the
unit simplex.  Let $\cal{T}$ be the set of all such points $u$. In the
usual way, the support of each element of $\cal{T}$ specifies a set of
elements of $X$.

Let $Y = T(X)$, and have $S(Y) = {\cal{R}}^Y$. Have $g$ be a uniform
average of a countable set of delta functions, each of which is
centered on a member of $\cal{T}$. So each of the delta functions
making up $g$ specifies a set of elements of $X$; $g$ is a
specification of a collection of such $X$-sets.

In this scenario $\sigma(E_g(y))$ is $\sigma$ applied to the union
(over $X$) of all the $X$-sets specified in $g$. In contrast,
$E_g(\sigma)$ is the average value you get when you apply $\sigma$ to
one of the $X$-sets specified in $g$. ${\cal{C}}\sigma(g)$ is the
difference between these two values. Intuitively, it reflects how much
overlap there is among the $X$-sets specified in $g$.

$ $

 \noindent {\bf{Example 11}}: Say $X$ has $N$ values, with $T(X) =
{\cal{R}}^X$.  Have $Y = T(X)$, and $S(Y)= {\cal{R}}^Y$, i.e., the set
of all nowhere-negative non-zero functions over those points in
${\mathbb{R}}^N$ with no negative components.  Choose $\sigma(y) =
H({y}) \; \forall y \in Y$, where $H(.) = -\int dy \; y(x)
{\mbox{ln}}[y(x)]$, the Shannon entropy function extended to
non-normalized $y$. This $\sigma$ is a natural choice to measure how
``spread out'' any point in $Y$ with magnitude 1 is. 

Have $g$ be a sum of a set of delta functions, about the distributions
over $\mathbb{B}$, $\{v^1, v^2, \ldots\}$. Then ${\cal{C}}\sigma(g)$
is a measure of how ``spread out'' those distributions are. In the
special case where $g = \delta_{v^1} + \delta_{v^2}$,
${\cal{C}}\sigma(g)$ is the Jensen-Shannon divergence between $v^1$
and $v^2$ \cite{tops02,lin91}. More generally, if $g$ is a probability density
function across the space of all distributions over $\mathbb{B}$,
${\cal{C}}\sigma(g)$ is a measure of how ``spread out'' that density
function is. 

$ $

There are several elementary properties of concavity gaps worth
mentioning: 

$ $

 \noindent {\bf{Lemma 2}}: 

 \begin{enumerate}
 \item $\cal{C}$ is linear.

 \item $\cal{C}\sigma$ is linear $\Leftrightarrow$ it equals 0
everywhere $\Leftrightarrow$ $\sigma$ is linear.

 \item $\cal{C}\sigma$ is continuous $\Leftrightarrow$ $\sigma$ is
continuous.

 \item ${\cal{C}}\sigma(g) = 0$ if $g \propto \delta_{y'}$ for some $y'
\in Y$.

 \item Giving $\cal{C}\sigma$ and the values of $\sigma$ at $1 + |Y|$
  distinct points in $Y$ fixes the value of $\sigma$ across all $Y$.
  ($|Y|$ is the dimension of $Y$.)

 \item The equivalence class of all $\sigma'$ having a particular 
concavity gap $\cal{C}\sigma$ is the set of functions of $y \in Y$
having the form $\{\sigma(y) + b \cdot y + a \;:\; a \in {\mathbb{R}},
b \in Y, \sigma(y) + b \cdot y + a \in S(Y)\}$.

 \end{enumerate}

$ $

{\bf{Proof}}: (1.1) and (1.4) are immediate. The first iff in (1.2)
follows from the fact that ${\cal{C}}\sigma(g) =
{\cal{C}}\sigma(\alpha g) \; \forall \alpha \in \mathbb{R}$. To see
the forward direction of the second iff, take $g = \delta_{y}/2$ and
$h = \delta_{y'}/2$ and expand ${\cal{C}}\sigma(g + h) = \sigma(y +
y') - [\sigma(y) + \sigma(y')]$. To see the forward direction of
(1.3), choose $g$ to have its center of mass infinitesimally to one
side of the discontinuity in $\sigma$, and then move it
infinitesimally to the other side to get a discontinuity in the
associated values of $({\cal{C}}(\sigma))(g)$.

 \noindent To prove (1.5), consider the case where $Y$ is
one-dimensional for simplicity. Say I give you $\sigma$ at $A$ and at
$B > A$, and also give you ${\cal{C}}\sigma$. Then for every $C > B$,
choose $g = \frac{B-C}{A-C} \delta_A + \frac{A-B}{A-C}
\delta_C$. Evaluating the associated value ${\cal{C}}\sigma(g) =
\sigma(B) - \frac{B-C}{A-C}\sigma(A) - \frac{A-B}{A-C}\sigma(C)$
allows us to solve for $\sigma(C)$. Similar reasoning holds for $C <
A$ and $C \in (A, B)$. For higher dimensions we need the value of
$\sigma$ at one extra point for each extra dimension of $Y$. This
completes the proof.

 \noindent To prove (1.6), first note that all members of that set do
indeed have the same concavity gap, $\cal{C}\sigma$.  To complete the
proof we must show that there are no other $\sigma$ with that
concavity gap. Let $\sigma'$ be any element of $S(Y)$ with the same
concavity gap as $\sigma$. By (1.5), if we know the value of $\sigma'$
at a total of $1 + |Y|$ points in $Y$, then we know $\sigma'$ {\it{in
toto}}. In turn, for any such set of $1 + |Y|$ values, we can always
find an $a$ and $b$ such that $a + b \cdot y + \sigma(y)$ lies in
$S(Y)$ and has those values. This means that $\sigma'$ is identical to
that $a + b \cdot y + \sigma(y)$. {\bf{QED.}}

$ $

 \noindent By (1.4), $\cal{C}\sigma$ necessarily obeys the second part
of condition (2) if $S(Y) = {\cal{R}}^Y$.

Next define a (strict) {\bf{dispersion}} over a space $X$ as a
(strictly) concave real-valued function over ${\cal{R}}^X$ that obeys
conditions (2) and (3) of a multimetric $\forall u, v, w \in
{\cal{R}}^X$.

$ $

 \noindent {\bf{Example 12}}: Take $X = \{1, 2\}$, with $T(X) =
{\cal{R}}^X = {\mathbb{R}}^2 - \{0\}$. Define $\sigma(u \in
{\mathbb{R}}^2)$ to equal 0 if $u_1 = 0$ or $u_2 = 0$, and equal
${\mbox{ln}}(1 + u_1) + {\mbox{ln}}(1 + u_2)$ otherwise. Then $\sigma$
is a (not everywhere continuous) strict dispersion.

$ $ 

 \noindent {\bf{Example 13}}: The $X, {\cal{R}}^X$, and $\rho$ of Ex. 3 
form a strict dispersion (see appendix).

$ $

 \noindent {\bf{Example 14}}: The $X, {\cal{R}}^X$, and $\sigma$ of
Ex. 5 form a dispersion.

$ $

 \noindent {\bf{Example 15}}: The $X, {\cal{R}}^X$, and $\sigma$ of
Ex. 11 form a strict dispersion.

$ $

There are several relations between concavity gaps and dispersions:

$ $

 \noindent {\bf{Lemma 3}}: Let $T(X) = {\cal{R}}^X$.

 \begin{enumerate}
 \item $\sigma$ is a dispersion over $T(X)$ $\Rightarrow$
$\sigma$ is nowhere-decreasing over $T(X)$.

 \item $\sigma$ is a dispersion over $T(X)$ and
$\sigma(s)$ is independent of $||s|| \; \forall \; s \ne 0 \in T(X)$
$\Rightarrow$ $\sigma$ is constant over the interior of $T(X)$.

 \item $\sigma$ is (strictly) concave over $V(X)$ $\Leftrightarrow$ 
$\cal{C}\sigma$ obeys condition (2) in full (and condition (3)) over $T(X)$.

 \item Say that $\sigma$ is continuous over $T(X)$. Then
$\cal{C}\sigma$ is separately (strictly) concave over each simplex in $T(X)$
$\Leftrightarrow$ $\sigma$ is (strictly) concave over $T(X)$.

 \end{enumerate}

$ $

 \noindent {\bf{Proof}}: (3.1) arises from the fact that a dispersion
is both concave and nowhere negative. To establish (3.2), first
consider any two vectors $u, v$ in the interior of $T(X)$ that differ
in only one component, $i$. Since no component of $u$ equals 0, there
must be an $s \in T(X)$ such that $\frac{u}{||u||} = \frac{v +
s}{||v+s||}$. (If $u_i > v_i, s = (u_i - v_i) \delta_i$. Otherwise
$s_i = 0$, and $s_j = u_j[\frac{v_i}{u_i} - 1] \; \forall j \ne i$.)
component is $i$, which is set so that $\frac{v_i + s_i}{v_j + s_j} =
\frac{u_i}{u_j}$ for every $j \ne i$.) By (3.1), this means that if
$\sigma$ is independent of the magnitude of its argument, $\sigma(v)
\le \sigma(u)$. Since the reverse argument must also hold, we have
$\sigma(u) = \sigma(v)$. Now repeat this reasoning to equate
$\sigma(v)$ with $\sigma(w)$ for some $w$ that differs from $v$ in
only one component, but differs from $u$ in two components. Continuing
in this way, we equate $\sigma(u)$ with $\sigma(z)$ for any $z$ that
differs in an arbitrary number of components from $u$.

 \noindent (3.3) is immediate from the definition of concavity and
Jensen's inequality.  To derive (3.4), first expand
${\cal{C}\sigma}(\frac{a+b}{2}) - \frac{{\cal{C}}\sigma(a) +
{\cal{C}}\sigma(b)}{2} \; = \; \sigma(\frac{E_a(y) + E_b(y)}{2}) -
\frac{\sigma(E_a(y)) + \sigma(E_b(y))}{2}$ when $||a|| = ||b||$.
($y$ being a generic argument of $T(X)$.) Next invoke (1.3) and
Jensen's inequality. {\bf{QED.}}

$ $

Let $f: {\mathbb{R}} \rightarrow {\mathbb{R}}$ be monotonically
increasing and strictly concave. Then by Lemma 3.3, if $\sigma$ is
strictly concave, $f({\cal{C}}\sigma)$ obeys conditions (2) and
(3). For example, this is the case for $\sqrt{{\cal{C}}\sigma}$. In
other words, so long as $\sigma$ is a strict dispersion,
$\sqrt{{\cal{C}}\sigma}$ obeys those conditions.

On the other hand, Lemma 3.2 means that any nontrivial $\sigma$ that
normalizes its argument (so that it is a probability distribution) and
then evaluates a function of that normalized argument cannot be a
dispersion.  So for example, if a concavity gap is a dispersion, it
must be constant.

Fortunately it is not the case that if $f({\cal{C}}\sigma)$ is a
multimetric it must be constant. In particular, often for a strictly
concave $\sigma$, $\sqrt{{\cal{C}}\sigma}$ for space $\{Y, S(Y)\}$ is a
multimetric for an appropriate $T(Y) \subseteq S(Y)$. 

$ $

 \noindent {\bf{Example 16}}: Choose $\{\sigma, Y, S(Y)\}$ as in
Ex. 11, and take $T(Y)$ to be all elements of $S(Y)$ which are sums of
two delta functions. This $\sigma$ is strictly concave, so we know
conditions (2) and (3) are obeyed by $\sqrt{{\cal{C}}\sigma}$.
Furthermore, for this choice of $T(Y)$, obeying condition (1) reduces
to obeying the conventional triangle inequality of two-argument
metrics, and it is known that the square root of the Jensen Shannon
divergence obeys that inequality ~\cite{tops02,futo04}. Therefore all
three conditions are met.

$ $

 \noindent {\bf{Example 17}}: Choose $\{\sigma, Y, S(Y)\}$ as in
Ex. 9. As in Ex. 16, this $\sigma$ is strictly concave, and therefore
$\sqrt{{\cal{C}}\sigma}$ automatically obeys conditions (2) and
(3). Now take $T(Y) = S(Y)$. Write ${\cal{C}}\sigma(g)$ as $\langle g,
g \rangle$ for the tensor contraction of Ex. 3, where $F(y, y') =
\frac{(y - y') \cdot (y - y')}{2}$. So by that example, we know that
$\sqrt{{\cal{C}}\sigma}$ is a multimetric.

 \section{Potential uses of multimetrics}

In addition to their intrinsic mathematical interest, multimetrics
have numerous potential applications. One of them is to allow more
nuanced complexity measures for physical systems, as described in
~\cite{woma04a}. Following is a list of some other from machine
learning~\cite{duha00}:

 \begin{enumerate}

 \item Mixture of Gaussians density estimation: In density estimation
one is given a data set of vectors $\{x^i\}$ that were generated by
IID sampling an unknown distribution over $X$ and wants to infer that
distribution. Say we have a probability distribution across $X$ given
by a linear combination of $n$ Gaussian distributions over $X$,
centered at the $n$ points $\mu^i$. Such a distribution induces a
probability of the set $\{x^i\}$. Accordingly, one way to estimate the
distribution that generated $\{x^i\}$ is to search for the linear
combination of $n$ Gaussians that maximizes the associated probability
of $\{x^i\}$.

One shortcoming of this procedure is that even though there are a
total of $n$ points used to parameterize the distribution, that
distribution is based solely on metric values for pairs of points
(namely the distances between $x$ and each of the $\mu^i$).  If we
have a multimetric $\rho$ though, we have several ways to avoid
this. For example, we could model the probability of each $x^i$ as a
Gaussian of $\rho(\delta_{x^i} + \sum_j \delta_{\mu^j})$.  We would
then take the probability of $\{x^i\}$ to be the product of the
probabilities of the $x^i$, as in conventional Gaussian mixtures
modeling. We could even model the probability of $\{x^i\}$ given the
$n$ points $\mu^i$ as a single Gaussian, with argument 
$\rho(\sum_i \delta_{x^i} + \sum_j \delta_{\mu^j})$.

 \item Kernel density estimation: In kernel density estimation, one
does not estimate the distribution over $x$ as a linear combination of
$n$ kernel functions (e.g., Gaussians) that are free to be centered
anywhere in $X$, and then search for which such linear combination
maximizes the probability of one's data.  Instead one centers a kernel
function on each of the data points, and searches for the optimal
parameters of those kernels functions. Conventionally such kernel
functions only take two arguments. However exactly as in application
1, if one has a multimetric over $X$, one can use kernel functions
whose argument involves more than two points at once.

 \item Classification can always be done via density estimation and
Bayes' theorem. So with applications 1 and 2, we have new ways of doing
classification.

 \item Kernel machines are a recent advance in machine learning in
which data is first mapped non-linearly into a feature space where
standard algorithms (like linear regression, linear discriminant
analysis, PCA, etc) are applied \cite{chsh00}. Because of the
non-linear mapping such methods work even when relationships in the
data are highly non-linear. All that is required for such methods is a
positive definite kernel function, $k(x,x')$, giving inner products in
the feature space. Multimetrics are not positive definite functions
but can easily be made so by taking $k(x,x') = \exp[-\rho(\delta_x +
\delta_{x'})]$ as the kernel. So any of the multimetrics discussed
above can be used for statistical analysis with kernel-based learning
algorithms. In particular, this is the case for either for supervised
or unsupervised learning with kernel machines. In particular, we can
use exponentials of multimetrics for regression by using them instead
of the conventional kernels of kernel machines, with the
multiplicative coefficients of each kernel (in the linear combination
that gives our fit to the data) set to minimize some appropriate
quadratic objective function.

 \end{enumerate}

 \section{Appendix}

 \subsection{Proof of Lemma 1.3}

That $f(\rho)$ obeys conditions (2) and (3) when $\rho$ does is
immediate. To prove that condition (1) is obeyed, consider any $u, v,
w \in T(X)$ such that $\rho(u + v) \le \rho(u + w) + \rho(v +
w)$. First assume that $\rho(u + v) \le {\mbox{max}}[\rho(u + w),
\rho(v + w)]$. Then since $f$ is increasing, $f(\rho(u + v)) \le
{\mbox{max}}[f(\rho(u + w)), f(\rho(v + w))]$. Since in turn
${\mbox{max}}[f(\rho(u + w)), f(\rho(v + w))] \le f(\rho(u + w)) +
f(\rho(v + w)$, condition (1) is obeyed. 

Now consider the other case, where $\rho(u + v) > {\mbox{max}}[\rho(u
+ w), \rho(v + w)]$. In this situation, because $f$ is concave, we
know that $f$ increases $\rho(u + v)$ less than it increases both
$\rho(u + w)$ and/or $\rho(v + w)$. So again condition (1) is
obeyed. {\bf{QED}}.

 \subsection {Proof of claim in Ex. 1}
Consider any $u, v$ and $w$ whose norms equal 1.  Then
squaring both sides of condition (1) for our $\rho$ implies that

$Var(\frac{u + v}{2}) \le Var(\frac{u + w}{2}) + Var(\frac{v + w}{2})
+ 2 \sqrt{Var(\frac{u + w}{2})Var(\frac{v + w}{2})}$.

 \noindent Use the expansion $Var(\frac{s + t}{2}) = \frac{Var(s) +
Var(t)}{2} + (\frac{E_s(x) - E_t(x)}{2})^2$ and cancel terms. The
hardest case for the resultant inequality to hold is where our three
variances all equal 0. Setting them to 0, we see that condition (1)
holds if for any three real numbers $a, b, c$,

$|a - b| \le |a - c| + |b - c|.$

 \noindent This is just the conventional triangle inequality though. So
condition (1) always holds. {\bf{QED.}}

 \subsection {Proof of claim in Ex. 3}
First note that for any $s \in T(X), {\langle}s \mid s{\rangle} \ge
0$, since all all components of $F$ are non-negative. Furthermore, all
$s, t \in T(X), {\langle}s \mid t{\rangle} \ge 0$, since all
components of those vectors are non-negative as are all components of
$F$. In addition, we can use the properties of $F$ to prove that our
tensor contraction obeys the Cauchy-Schwartz inequality: ${\langle}u
\mid v{\rangle}^2 \le {\langle}u \mid u{\rangle}{\langle}v \mid
v{\rangle} \; \forall u, v \in T(X)$. (Exampnd $\langle s, s \rangle
\ge 0$ for $s \equiv u - \alpha v$. Solve for the $\alpha$ minimizing
the lefthand side (which is quadratic in $\alpha$), and plug that
in. Collecting terms establishes the desired inequality.)

Now to check condition (1) for our $\rho$, square both sides of it and
cancel terms. So the lefthand side is just ${\langle}u \mid
v{\rangle}$. Since all inner products are non-negative, the right-hand
side is bounded below by $\sqrt{{\langle}u \mid u{\rangle}{\langle}v
\mid v{\rangle}}$. Plugging in the Cauchy-Schwarz inequality
establishes that condition (1) does indeed hold. {\bf{QED.}}

 \subsection {Proof of claim in Ex. 4}
Conditions (2) and (3) are immediate. To prove condition (1), first
note that it holds for $\rho_1(s) = {\mbox{max}}(x_1 : s(x_1 \ne
0)$. Then note that it holds for $\rho_1(s) = -{\mbox{min}}(x_1 :
s(x_1 \ne 0)$, and invoke Lemma 1.2, to see that the width in $x_1$
of the support obeys condition (1). {\bf{QED.}}

 \subsection {Proof of claim in Ex. 5}
Conditions (2) and (3) are immediate. Condition (1) also holds if (the
supports of) $u$ and $v$ overlap, since any non-zero volume must equal
at least 1, and that overlap volume gets counted twice in the sum
$\rho(u + w) + \rho(v + w)$, regardless of $w$. If (the supports of)
$u$ and $v$ do not overlap, then (the support of) $w$ must either
extend outside of (the support) of $u$ or of $v$. This means that
condition (1) must hold in this case as well. {\bf{QED.}}

 \subsection{Proof of claim in Ex. 6}
Define ${\mbox{argmax}}_x u(x) \equiv a, {\mbox{argmax}}_x v(x) \equiv
b, {\mbox{max}}_x (u(x) + v(x)) \equiv M$, and ${\mbox{max}}_x (u(x) +
w(x)) + {\mbox{max}}_x (v(x) + w(x)) - 2||w|| \equiv N$; we want to
prove that $M > N$.  To that end, note that if the support of $w(x)$
is restricted to $a$ and $b$, then $N$ becomes $u(a) + v(b) - ||w|| =
u(a) + v(b) - 1 \le u(a)$. On the other hand, $M$ is bounded below by
$u(a)$. So condition (1) holds for this situation.

We now consider the situation where $w$'s support is not restricted to
$a$ and $b$. It will be useful to define ${\mbox{argmax}}_x (u(x) +
w(x)) = d$ and ${\mbox{argmax}}_x (v(x) + w(x)) = e$. First consider
the case where $d \ne e$. Then it is immediate that by transferring
any $w(c \not \in \{d, e\})$ to $w(d)$ and/or $w(e)$, we do not
decrease $N$ (since $||w||$ doesn't change). We can then transfer
$w(d)$ to $w(a)$ and $w(e)$ to $w(b)$, again not decreasing $N$.
After doing this for all such points $c$, we recover the case where
the support of $w(x)$ is restricted to $a$ and $b$, as in the
preceding paragraph. So we can conclude that condition (1) is obeyed
for this case.

The remaining case to consider is where $d = e$. For this case we can
transfer all $w(c \ne d)$ to $w(d)$, and in doing so increase
$N$. Doing this for all such $c$ restricts $w$'s support to $d$. After
having done this, ${\mbox{max}}_x (u(x) + w(x)) = u(d) + ||w|| \le
u(a) + 1$, and similarly ${\mbox{max}}_x (v(x) + w(x)) \le v(b) +
1$. So we again get $N \le u(a) + v(b) - 1$, which means $M \ge
N$. There are no more cases to consider. {\bf{QED.}}

 \subsection{Proof of claim in Ex. 13}
First note that we have already established that our $\rho$ obeys
conditions (2) and (3) of being a multimetric, and therefore only need
to establish that it is strictly concave. That will be the case iff
$\rho(\alpha u + (1 - \alpha)v) \le \alpha \rho(u) + (1 -
\alpha)\rho(v) \; \forall u \in T(X), v \in T(X), \alpha \in [0,
1]$. Square both sides of this inequality and cancel terms. Then
exploit the Cauchy Schwarz inequality for ${\langle}. \mid
.{\rangle}$, established in the proof of the claim in
Ex. 3. {\bf{QED.}}

 \section{Acknowledgements}

I would like to thank Bill Macready and Creon Levit for stimulating
discussion.


\begin{thebibliography}{10}

\bibitem{chsh00}
Nello Christianini and John Shawe-Taylor.
\newblock {\em An introduction to support vector machines and other
  kernel-based learning methods}.
\newblock Cambridge University Press, 2000.

\bibitem{duha00}
R.~O. Duda, P.~E. Hart, and D.~G. Stork.
\newblock {\em Pattern Classification (2nd ed.)}.
\newblock Wiley and Sons, 2000.

\bibitem{futo04}
Bent Fuglede and Flemming Topsoe.
\newblock Jensen-shannon divergence and hilbert space embedding.
\newblock Submitted to ISIT2004, 2004.

\bibitem{hall99}
M.~Hall.
\newblock Universal geometric approach to uncertainty, entropy, and
  information.
\newblock {\em Physical Review A}, 59:2602, 1999.

\bibitem{judg04}
George Judge.
\newblock Semiparametric moment based estimation for binary response models.
\newblock RMBE-DE1-4-2-04.doc, 2004.

\bibitem{lin91}
Jianhua Lin.
\newblock Divergence measures based on the shannon entropy.
\newblock {\em IEEE Trans. Info. Theory}, 37(1):145--151, 1991.

\bibitem{osva03}
Ferdinand Osterreicher and Igor Vajda.
\newblock A new class of metric divergiences on probability spaces and its
  applicability in statistics.
\newblock {\em Ann. Inst. Statist. Math.}, 55(3):639--653, 2003.

\bibitem{tops02}
Flemming Topsoe.
\newblock Inequalities for the jensen-shannon divergence.
\newblock 2002.
\newblock unpublished.

\bibitem{woma04a}
David~H. Wolpert and William Macready.
\newblock Self-dissimilarity as a high dimensional complexity measure.
\newblock In {\em Proceedings of the International Conference on Complex
  Systems, 2004}, 2004.
\newblock in press.

\bibitem{woma04b}
David~H. Wolpert and William Macready.
\newblock Self-dissimilarity as a high dimensional complexity measure.
\newblock http://xxx..., 2004.

\end{thebibliography}
 \end{document}